\documentclass[final,5p,times,twocolumn]{elsarticle}

\usepackage{graphicx}
\usepackage{amssymb}
\journal{International Journal of Food Properties}

\begin{document}

\begin{frontmatter}



\title{Quantifying thermally induced flowability of rennet cheese curds}

\author[Hadd]{Hiroyuki Shima}
\author[Madd]{Morimasa Tanimoto}
\address[Hadd]{Department of Environmental Sciences, University of Yamanashi, 4-4-37, Takeda, Kofu, Yamanashi 400-8510, Japan}
\address[Madd]{Department of Local Produce and Food Sciences, University of Yamanashi, 4-4-37, Takeda, Kofu, Yamanashi 400-8510, Japan}

\begin{abstract}
Conversion of liquid milk to cheese curds is the first stage in cheese manufacture. Changing the rigidity of cheese curds through heating and pH control is an established method for preparing fresh curds, whereas a similar method to prepare fully coagulated curds is largely unknown. This study elucidated the effect of temperature variation on the viscoelastic moduli of fully coagulated curds under different pH conditions. The results showed that rennet curds treated at pH 4.8 exhibited drastic changes in the viscoelasticity at 43 $^\circ$C, above which the degree of fluidity exceeded the degree of rigidity. The viscoelastic moduli exhibited exponential decay as a function of temperature, which was independent of pH.
\end{abstract}

\begin{keyword}
Casein micelle \sep Milk protein \sep Rennet clotting \sep Viscoelasticity \sep Dynamic shear modulus
\end{keyword}

\end{frontmatter}

% \linenumbers

\section{Introduction}
\label{}

Why does milk appear white? The answer is because of the presence of casein, a milk-specific protein comprising many nutritionally important amino acids. Casein is one of the best-known phosphorylated proteins, in which the phosphate groups are bound to many serine residues. The phosphate groups confer an overall negative charge to the casein molecule, enabling it to form a complex with Ca ions. The complexes are stabilized by assembling each other, resulting in numerous colloid particles (called casein micelles) that are uniformly suspended in the milk. In the presence of light, therefore, the casein micelles diffusively reflect incident light, because of which milk appears white ($i.e.$, by the Tyndall effect).

The casein in milk can be precipitated and separated by adding an enzyme called rennet to milk. The enzyme proteolyses the surface of casein micelles, causing them to attract each other. Consequently, numerous casein micelles aggregate, solidify, and finally precipitate. The rennet-induced aggregation of casein micelles has been utilized in the production of various casein products that serve as functional protein ingredients in diverse applications, including foods, pharmaceuticals, and cosmetics \cite{Korhonen,Semo,Livney,Shah}. Against this backdrop, understanding the essential properties of casein micelles and their aggregates is important both for the development of optimal dairy products and for the control of their functionalities \cite{Horne,Mellema2000,Esteves,Udabage,Srinivasan,Mellema2002}.

Structural transformation and development of firmness during milk clotting have been examined by sophisticated measurement techniques \cite{Lyndgaard,Tabayehnejad,Djaowe}. The structure and properties of casein micelles are highly dependent on the environmental conditions. In the production of natural cheese, for example, rennet-induced proteolysis and the resultant aggregation of casein micelles lead to cheese curd. The obtained cheese curd, reinforced by a three-dimensional casein network, exhibits complex viscoelasticity in accordance with changes in the temperature, pressure, pH, and protein concentration \cite{Nabulsi,Catarino}; similar complex variations have been observed in concentrated milk \cite{Morison} and mixed food proteins \cite{Onwulata}. Furthermore, it is envisaged that artificially heating and lowering the pH of the curds {\it after they have been fully coagulated} should cause the casein network to dissociate and rearrange, thus potentially increasing its fluidity. Most previous studies on this issue, nevertheless, have focused primarily on {\it fresh} curds obtained {\it just after} the initiation of casein micelle aggregation. Little attention has been paid to quantitative determination of thermally and/or pH-induced changes in the viscoelasticity of {\it already fully coagulated} curds, though the latter is crucial for the textural control of processed dairy products.

The present study is designed to investigate the effect of temperature variation on the viscoelastic moduli of fully coagulated rennet curds. Dynamic shear tests are performed to quantitatively assess the variation of the temperature dependence of the moduli with changes in pH. Molecular interpretations of viscoelastic behaviours are also presented.

\section{Sample Preparation}

The milk samples were purchased from dairy farms in the foothills of Mount Yatsugatake (on the border of Yamanashi and Nagano Prefectures), Japan. Milk clotting was performed by using a device, with a 20 L capacity, that is usually used for obtaining natural cheeses (Nichiraku Kikai Co., Ltd.) by following the procedure described below.

The milk samples were first pasteurized by maintaining 18 L of raw milk at 65 $^\circ$C for 30 min. This process eliminates the bacteria, thus preventing the degradation of milk proteins at high temperatures. The sample pH was 6.73 immediately after pasteurization. The sample was then cooled to 31 $^\circ$C, and 18 mL of a lactobacillus culture solution was added to it in order to acidify it. To prepare the culture solution, 0.3 g of Direct Vat Set (DVS) lactobacillus starter (CH-N11, Chr. Hansen, Nosawa \& Co., Ltd.) containing a blend of several types of concentrated, dried, and frozen lactobacilli was mixed with 300 mL of pasteurised milk, and the solution was cultured overnight at 21 $^\circ$C. After adding the culture solution, the sample was maintained at 31 $^\circ$C, which is the temperature that produces the greatest lactobacillus activity, for 30 min. The sample pH at this stage was 6.50.

Rennet was then added to the above sample that slightly acidified by lactobacilli. In this experiment, 0.5 g of rennet (CHY-MAX, Chr. Hansen, Nosawa \& Co., Ltd.) dissolved in sterile cold water was added to the sample, which was then left to stand at 31 $^\circ$C for 35 min. After the milk started to coagulate (pH 6.46), the sample was cut into cubes of 12-13 mm to remove the whey from the curds. Five minutes after cutting, the sample was gently agitated for 15 min to encourage the removal of the whey. As a result, whey corresponding to one-third of the original sample weight was eliminated, yielding fresh curd granules.

To completely eliminate the whey from the sample and adjust the pH to the target values (4.8-6.0), hot water (80 $^\circ$C) was added to the curd granules. The granules were then gently agitated at 0.2 $^\circ$C/min until a temperature of 38 $^\circ$C (pH 6.34) was achieved; eventually, all the whey was removed. In the final step, the curds were aged until they reached the target pH. The time required to attain curds with the lowest pH was approximately 6 h. After pH adjustment, a series of curds with different pH values were frozen and stored. Immediately before measurement, the curds were defrosted in a refrigerator and then stirred at 50-60 $^\circ$C.

\section{Dynamic Shear Measurement}

A thorough understanding of the viscoelastic properties of cheese curds is not only a major scientific challenge, but also of practical importance since many dairy products and processes rely on the control of these properties under an external load \cite{Taneya,Brown,Lucey2003}. Small-amplitude oscillatory shear analysis, a non-destructive protocol for determining the viscoelasticity of a material, is an efficient approach to address this issue \cite{Gunasekaran,Tunick}. In this analysis, oscillatory shear strain is applied to the sample while maintaining the strain in the linear viscoelastic region. Two main parameters are determined from this test: one is the elastic (or storage) modulus designated by $G'$, which is a measure of the elastic energy stored per oscillation cycle. Simply expressed, this parameter indicates the degree to which the sample gives a solid-like response to the dynamic load. The other parameter is the viscous (or loss) modulus designated as $G''$, which is a measure of the energy dissipated as heat per cycle, thus indicating the degree to which the sample shows liquid-like behaviour.

Actual measurements were performed using the Anton Paar MCR 302 rheometer. The samples were thinly sliced and sandwiched between two flat disk plates with 25 mm radius, facing each other, separated by a gap of 2 mm. The sample surface was coated with silicone oil to prevent evaporation of water during measurements. After coating the sample, it was gradually cooled from 50 $^\circ$C to 5 $^\circ$C at a rate of 2 $^\circ$C/min, during which the variations in $G'$ and $G''$ were measured by applying the oscillatory shear. For all the measurements, the angular frequency of oscillation was set to 1 Hz and the shear strain was 0.1\%. From the $G'$ and $G''$ data, the amplitude of the complex modulus $|G^*|=\sqrt{(G')^2+(G'')^2}$ was also evaluated under each temperature and pH condition.

\section{Results and Discussion}

%%%--------------------------------------------------------------
\begin{figure}[ttt]
\includegraphics[width=0.41\textwidth]{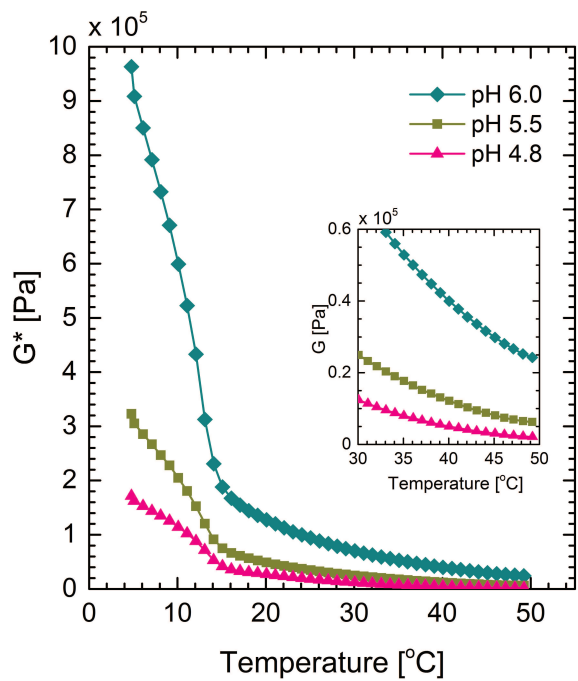}
\caption{Temperature dependence of the complex modulus amplitude ($|G^* |$) of fully coagulated curds under three different pH conditions. Inset: Enlarged view of the data in the high-temperature region.}
\label{fig01}
\end{figure}
%%%--------------------------------------------------------------

Figure \ref{fig01} shows the temperature ($T$) dependence of the complex modulus ($|G^*|$) of the fully coagulated curds under three different pH conditions: 4.8, 5.5, and 6.0. At each pH condition, 10 samples were analysed and only a slight sample dependence was evident from the curves of $|G^*|$. Each data point in Fig.~\ref{fig01} represents the mean value of $|G^*|$ for the 10 samples. From the figure, it is clear that for all the evaluated pH conditions, $|G^*|$ below 15 $^\circ$C exhibited a rapid decrease with increasing $T$, which changed to a slow decay till 50 $^\circ$C. The rapid decrease in $|G^*|$ in the low-$T$ region is attributed to the melting of lipid globules pooled in the sample. In fact, the curds produced herein were not delipidated, and thus contained a significant amount of lipid. These lipid globules solidify at low temperatures and are embedded into the voids of the casein network of the curds. Consequently, the entrained `solid' globules act as inert fillers that enhance the stiffness of the curds \cite{Vliet}. The effect of such fillers diminishes with increasing $T$ because of the softening of the solid globules. This accounts for the rapid decrease in $|G^*|$ that persists until T reaches the melting point of the lipid (≈15 $^\circ$C). It follows from Fig.~\ref{fig01} that even above the melting point, $|G^*|$ continues to decay slowly with increasing $T$. The slow decay indicates thermally excited flow of the molecules that constitute the casein network inside the curd.

\begin{figure}[ttt]
\includegraphics[width=0.48\textwidth]{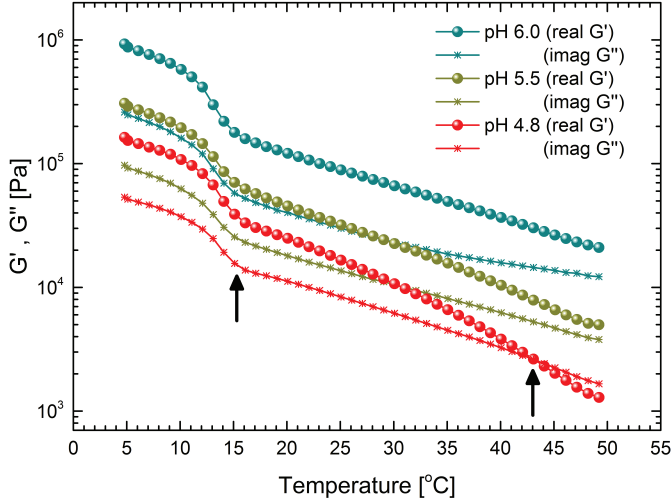}
\caption{Semi-logarithmic plot of the temperature variance of storage ($G'$) and loss ($G''$) moduli. Respective upward arrows indicate the sol-gel transition point (right) and the shoulder caused by lipid-globule melting (left).}
\label{fig02}
\end{figure}

Figure \ref{fig02} presents a single-logarithmic plot of the same data as in Fig.~\ref{fig01}, while the storage ($G'$) and loss ($G''$) components are plotted separately. The shoulder structures in the curves of $G'(T)$ as well as $G''(T)$ at 15 $^\circ$C become clear in these plots, as indicated by an upward arrow on the left. Interestingly, above 15 $^\circ$C, most of the curves show a nearly exponential decay with the functional form,
\begin{equation}
G'(T) = C \cdot \exp\left(-\frac{T}{A}\right) \label{eq01}
\end{equation}
with appropriate constants $C$ and $A$. For instance, above 15 $^\circ$C the decay of $G'(T)$ at pH 6.0 is represented by the form shown in Eq.~(\ref{eq01}) with $C=4.0\times 10^5$ Pa and $A=16.5$ $^\circ$C. Above 15 $^\circ$C, the other curves are fairly well fitted to the exponential form of Eq.~(\ref{eq01}), wherein the value of $A$ is insensitive to pH variation.

%%%--------------------------------------------------------------
%%%--------------------------------------------------------------

The most important observation in Fig.~\ref{fig02} is the occurrence of the sol-gel transition for the data acquired at pH 4.8. The transition point is highlighted by the right-hand-side arrow. In general viscoelastic materials, the structural transition from the sol phase (liquid state) to the gel phase (solid state) is signified by the intersection between the two curves of $G'(T)$ and $G''(T)$ \cite{Tung,RossMurphy}. Accordingly, the average critical temperature $T_{\rm C}$, at which the fully coagulated curds undergo the sol-gel transition, for 10 samples is estimated to be 43 $^\circ$C. This means that the sample at pH 4.8 exhibits solid-like behaviour below $T_{\rm C}$, but assumes viscous liquid behaviour above $T_{\rm C}$. The heat-induced flowability is believed to result from higher molecular mobility and reduced cross-linkage that are internal to the casein network \cite{Mleko2005}. These two (and maybe other) physicochemical factors promote the molecular alignment in the direction parallel to the tensile direction, enhancing the flow of the cheese curds at $T>T_{\rm C}$. The results also confirmed that the value of $T_{\rm C}$ is unique and reproducible for an identical sample, though there is a slight sample-dependent fluctuation of the order of a few degrees Celsius.

The structural transition from liquid to solid (and vice versa) demonstrated in Fig.~\ref{fig02} is consistent with the thermal softening of the casein network in fully coagulated Mozzarella cheese reported in the literature \cite{Ak1995}. Using the squeezing flow method, it has been reported that Mozzarella cheese shows a decreased resistance to flow with increasing temperature; the relaxation time of the Mozzarella cheese was reduced by several times on increasing the temperature from 30 $^\circ$C to 60 $^\circ$C, similar to the present observation for $|G^*|$ (see inset of Fig.~\ref{fig01}). Compared to a previous work \cite{Ak1995}, the present work offers the distinction that the characteristic temperature $T_{\rm C}$ for the transition can be successfully determined in an objective and quantitative manner by oscillatory shear analysis.

It should be stressed that the physicochemical origin of the exponential decay of $G'$ and $G''$ is yet to be confirmed. To our knowledge, however, there is a lacuna in the understanding of how casein micelles and other chemical components interact with each other in fully coagulated curds and how these interactions evolve the viscoelastic properties; this situation is in contrast with the now reasonably well-established structure and composition of individual casein micelles before aggregation \cite{Fox,Dalgleish2011,Dalgleish2012,Kruif,Mezzenga2013}. It is thus conjectured that a theoretical description of the exponential decay represented by Eq.~(\ref{eq01}), if successfully developed, should shed light on the mechanism of the interaction of cheese curd constituents; this subject will be addressed in our future work.

Before closing the article, we would like to point out that a certain class of cheese demonstrates thermally induced solidification, opposite to the present observation. This cheese is called $paneer$ \cite{Khan}, a South Asian variety of soft cheese prepared by acid coagulation. The ability of paneer to be deep fried ($i.e.$, non-melting property even at high temperatures) is in contrast with the enhanced melting of rennet cheeses. Microscopic interpretation of the contrasting thermal responses in the viscoelasticity of paneer (acid-based cheese) relative to rennet cheeses has not been established, thus remaining a challenge in the field of dairy science.

\section{Conclusion}

The sol-gel phase transition of fully coagulated curds driven by temperature and pH controls was quantified herein with the help of small amplitude oscillatory shear measurements. For the curds analysed, the transition occurs at 43 $^\circ$C when the samples are subjected to pH 4.8. The occurrence of phase transition in fully coagulated curds implies a novel technique for the manipulation of the cheese texture through pH regulation and heat treatment, which will be complementary to the well-established technique for preparing fresh curds. The results also show that the viscoelastic moduli at different pH conditions universally follow an exponential decay as a function of temperature. This finding may provide a platform for better understanding of the structural relaxation and molecular interaction mechanism of the casein network within fully coagulated curds.

\section*{Acknowledgement}

The authors cordially acknowledge the fruitful discussions held with Prof.~Ryoya Niki and Prof.~Katsuyoshi Nishinari. Gratitude is also expressed to Ms.~Tazuko Watanabe (Anton Paar Japan K.K.), Mr.~Toshiaki Shioya (Unitec Foods Co., Ltd.), and Mr.~Kunio Ueda (Yueisha) for their technical support with measurements and sample preparation. This work was supported by JSPS KAKENHI Grant Numbers 25390147 and 25560035. H.S. gratefully acknowledges financial support from the Kieikai Research Foundation.



%% References


%% \bibliographystyle{elsarticle-num}
%% \bibliography{<your-bib-database>}



\end{document}